\documentstyle[prl,twocolumn,aps]{revtex}
\tightenlines
\RequirePackage{epsfig}
\begin{document}
\draft
\title{Generation of electromagnetic pulses from plasma channels
induced by\\femtosecond light strings}
\author{Chung-Chieh Cheng, E. M. Wright, and J. V. Moloney\\
Arizona Center for Mathematical Sciences, and Optical Sciences
Center, Tucson, AZ 85721, USA}
\date{\today}
\tolerance = 1000
\maketitle
\begin{abstract}
We present a model that elucidates the physics underlying the
generation of an electromagnetic pulse from a femtosecond laser
induced plasma channel. The radiation pressure force from the
laser pulse spatially separates the ionized electrons from the
heavier ions and the induced dipole moment subsequently oscillates
at the plasma frequency and radiates an electromagnetic pulse.
\end{abstract}
Recent investigations of the propagation of intense femtosecond
infrared (IR) laser pulses in air show that the dynamical
interaction between nonlinear self-focusing, plasma defocusing,
and group-velocity dispersion can cause an initial beam to break
up spatially into several filaments, or light strings, with
diameters around a hundred microns that can maintain themselves
over long distances
\cite{Braun:1995,Nibbering:1996,Brodeur:1997,Mlejnek:1998,Mlejnek:1999}.
It has been observed experimentally that femtosecond light strings
in turn produce plasma channels by multi-photon ionization (MPI)
along their direction of propagation with lengths on the order of
centimeters
\cite{Schillinger:99,Tzortzakis:99,Tzortzakis:2000,Vid00,Proulx:2000}.
Observations of the electromagnetic pulses (EMPs) from light
string induced plasmas suggest that these channels attain dipole
moments during the laser pulse which subsequently oscillate at the
plasma frequency and radiate \cite{Proulx:2000}.

In this letter we present a simple model for EMP generation from
femtosecond light string induced plasmas to elucidate the
underlying physics, in particular how the initial dipole moment is
produced in the plasma channel, and the characteristic field
strengths and spectrum of the EMP. In this model electrons and
ions are initially created by MPI due to the light string, and a
dipole moment is subsequently induced in the plasma via the
radiation pressure force resulting from the velocity-dependent
Lorentz force. For a laser field of frequency $\omega$ this force
contains a second-harmonic component at $2\omega$ in addition to a
low frequency component, and it is this low frequency component of
the force that creates the initial dipole moment in the plasma
channel by longitudinally separating the light electrons from the
heavy ions on the short time scale of the light string. Following
the laser pulse the electron-ion system then starts to oscillate
longitudinally at the plasma frequency \cite{TonLan29} and
generates an EMP propagating radially away from the plasma channel
\cite{Proulx:2000}. We stress that this mechanism for creating a
dipole moment in a plasma is new and distinct from that due to the
ponderomotive force from a tightly focused Gaussian beam
\cite{Hamster:1993,Goreslavsky:1995}: In particular, the usual
ponderomotive force acting on electrons is proportional to the
spatial gradient of the applied light intensity \cite{LL}. For
light strings of diameter $100$ $\mu$m and a few centimeters
length the ponderomotive force should be negligible, and this was
our motivation for looking for a new physical process that could
create a dipole moment in the plasma channel.

In our model we consider a femtosecond laser pulse linearly
polarized along the x-axis, and propagating in air along the
positive z-axis, with a central frequency $\omega$ in the IR
region. For an electron plasma in the presence of an electric
field $\vec{E}$ and a magnetic field $\vec{B}$, the hydrodynamic
equation for the average velocity of an electron $\vec{v}$ reads
\cite{Bloembergen:1968}
\begin{equation}
\frac{\partial\vec{v}}{\partial t} + \,\,({\rm \vec {v}} \cdot
\nabla ){\rm \vec {v}} = \,\, - {\frac{{e}}{{m}}}\left( {\vec
{E}+\eta\vec{v}}\times\vec {B} \right)\, +
\frac{\partial\vec{v}}{\partial t}|_{scatt} \quad {\rm ,}
\label{veq}
\end{equation}
where $e$ and $m$ are the charge and mass of the electron, and
$(\partial\vec{v}/\partial t)|_{scatt}$ is the time rate of change
of the velocity due to constituent particle scattering. The
parameter $\eta$ is included to keep track of orders of
perturbation theory and reflects the fact that the
velocity-dependent Lorentz force is order $|\vec{v}|/c<<1$ smaller
than the dominant force due to the electric field under the
non-relativistic conditions appropriate here. To keep the theory
as simple as possible we concentrate on the electron motion,
assuming that the heavy ions remain stationary. In addition, we
have the continuity equation for the electron density
\cite{Mlejnek:1998,Bloembergen:1968}
\begin{equation}
\frac{\partial N_e}{\partial t}+ \,\,\nabla \cdot \left( {N_{e}
\,{\rm \vec {v}}} \right) = \sum_K{\frac{{\beta ^{(K)}}}{{K\,\hbar
\,\omega} }} I^K(\vec {r},t) {\rm ,} \label{Neq}
\end{equation}
where $N_e$ is the electron density, $\omega$ is the central
frequency of the input laser pulse, $K$ is the order of the MPI
and the sum allows for the various constituents of air, mainly
oxygen and nitrogen molecules, $\beta^{(K)}$ is the MPI
coefficient derived from, for example, Keldysh theory
\cite{Keldysh:1965}, and $I(\vec{r},t)$ is the intensity
distribution of the laser pulse. The right-hand side of Eq.
(\ref{Neq}) describes the generation of electrons via MPI, since
the effects of avalanche ionization for femtosecond pulses are
negligible compared to MPI. To complete the system we couple these
equations to the Maxwell equations for the electromagnetic field
incorporating the effects of nonlinear self-focusing, plasma
defocusing, and group velocity dispersion. The key Maxwell
equation we need here is
\begin{equation}
\label{divE} \nabla\cdot\vec {E}=-e(N_{e}-N_{i})/\epsilon_0 ,
\end{equation}
which determines the irrotational part of the electric field from
the net charge density $-e(N_e-N_i)/\epsilon_0$, with $N_i$ the
ion density of the plasma.

Equations (\ref{veq}-\ref{divE}) provide the basic dynamics of the
plasma. Detailed numerical simulations of the formation and
structure of light strings have been performed in a variety of
situations
\cite{Braun:1995,Nibbering:1996,Brodeur:1997,Mlejnek:1998,Mlejnek:1999}.
Typically, light strings have a diameter of around 100 $\mu$m and
a temporal duration $t_p\approx 10$ fs. Due to temporal
compression associated with self-focusing of the input pulse
\cite{Marb}, the duration $t_p$ of the light string is much
shorter than the input pulse duration. Here we shall use electric
and magnetic fields which reflect the characteristic properties of
a single light string rather than explicitly solving for their
detailed structure, which would obscure the physics to be
presented.

To proceed we first consider the evolution of the electron
velocity for short times on the order of the light string duration
$t_p$ during which time the electrons are generated. For such
short times we may safely neglect the effects of particle
diffusion, hence we ignore spatial derivatives in Eq. (\ref{veq}).
When the electrons are liberated via MPI they are accelerated in
the laser field and undergo collisions, and we approximate the
scattering as $(\partial\vec{v}/\partial t)|_{scatt}\approx
-\gamma\vec{v}$, with $\gamma\approx 10^{12}-10^{13}$ s$^{-1}$ the
collision rate for air at atmospheric pressure
\cite{Proulx:2000,Ready}. Furthermore, the dominant
electromagnetic fields will be those due to the light string
$\vec{E}=\vec{E}_L$ and $\vec{B}=\vec{B}_L$. The spatial
dimensions of the light string and plasma channel are around 100
$\mu$m$^2\times 1$ cm, enormous on the scale of a cubic
wavelength, and we approximate the fields as spatially uniform
over the transverse profile of the plasma channel. The fields
$\vec{E}_L$ and $\vec{B}_L$ due to the light string can then be
expressed as
\begin{equation}
\vec{E}_L(\vec{r},t) = \frac{\hat x}{2} \left({\cal E
}_L(t-z/v_g){e^{i(kz-\omega t)} + c.c.} \right) , \label{EB}
\end{equation}
with $\vec {B}_L(\vec{r},t) = \hat{y}E_L/c$, where ${\cal
E}_L(\tau)=\sqrt{2I_{0}/c\epsilon_{0}}\cdot e^{-\tau^2/t_p^2}$ is
the Gaussian approximation to the light string electric field
envelope with $\tau=t-z/v_g$ the retarded time in a reference
frame moving at the light string group velocity $v_g$, $I_{0}$ and
$t_p$ being the peak intensity and duration of the light string,
$n_0\approx 1$ is the refractive-index of air, and $k=\omega/c$ is
the magnitude of the optical wave vector. Since the light string
is polarized along the x-direction we expect the electron motion
to be predominantly along the x-axis, and we correspondingly write
the electron velocity as $\vec{v}=\hat{x}v_x+\eta\hat{z}v_z$: By
symmetry the electron motion will be confined to the x-z plane.
Then to zeroth and first orders in the perturbation parameter
$\eta$ Eq. (\ref{veq}) yields the pair of equations
\begin{eqnarray}
\frac{\partial v_x}{\partial t} &=& -\frac{e}{m}E_L - \gamma v_x ,
\nonumber \\ \frac{\partial v_z}{\partial t} &=& -\frac{e}{m}\left
(\frac{v_x}{c}\right )E_L - \gamma v_z  . \label{vxz}
\end{eqnarray}
Under the assumption that the electric field envelope ${\cal
E}_L(\tau)$ varies slowly on the carrier time scale $\omega^{-1}$
the equation for $v_x$ may be solved approximately as
\begin{equation}
v_x(z,t)\approx -\frac{e}{2m} \left (\frac{ {\cal
E}_L(t-z/v_g)e^{i(kz-\omega t)} }{(\gamma-i\omega)} + c.c \right )
\label{vxapp}.
\end{equation}
Next, inspection of the driving force term proportional to
$-e(v_x/c)E_L$ appearing in Eq. (\ref{vxz}) for $v_z$ reveals that
it has high frequency or second-harmonic terms varying as $\pm
2\omega$ in addition to a low frequency term. The low frequency
term produces a force that is always directed along the positive
z-axis, and creates a charge separation between the fast
responding electrons and slow ions along the direction of the
light string, and hence an initial dipole moment. In contrast, the
charge separation induced by the second-harmonic term keeps
switching direction, hence producing zero dipole moment on
average, and we therefore isolate the low frequency component.
Then by substituting Eqs. (\ref{EB}) and (\ref{vxapp}) into
(\ref{vxz}) for $v_z$ we obtain
\begin{eqnarray}
\frac{\partial^2\xi}{\partial t^2}&+&\gamma\frac{\partial
\xi}{\partial t} = \frac{F_{RP}}{m} \nonumber \\ F_{RP}(z,t) &=&
\frac{e^2}{2mc}\frac{\gamma}{(\gamma^2+\omega^2)}|{\cal
E}_L(t-z/v_g)|^2 , \label{zetaq}
\end{eqnarray}
where $v_z=\partial\xi/\partial t$, $\xi(z,t)$ being the
longitudinal $(z)$ electron-ion separation, and $F_{RP}$ is the
radiation pressure force exerted on the electrons (see below).
Integrating Eq. (\ref{zetaq}) over time and using
$\xi(z,-\infty)=0$ and $\partial\xi/\partial t|_{\pm\infty}=0$ we
obtain
\begin{equation}
\xi_0 = \frac{e^2}{\epsilon_0m^2\omega^2c^2}
\int_{-\infty}^{\infty} I(\tau)d\tau , \label{xi0}
\end{equation}
where $I=\frac{1}{2}\epsilon_0c|{\cal E}_L|^2$ is the light string
intensity, we have used $\omega>>\gamma$ to simplify the
expression, and we have extended the upper time limit to infinity
with the understanding that we formally only need a large enough
time window to cover the light string duration.

The quantity $\xi_0$ is the longitudinal electron-ion separation
induced by the light string, with associated dipole moment
directed along the z-axis $d_0=-e\xi_0$, and is proportional to
the time-integrated intensity or fluence of the light string, but
relatively insensitive to the scattering rate so long as
$\omega>>\gamma >0$. In general, $\xi_0$ could depend on position
along the light string due to longitudinal variation of the
intensity profile, but in keeping with our approximations we shall
treat $\xi_0$ as a constant over the length of the plasma channel.
Physically, the low frequency component of the Lorentz force
exerts a force on each electron in the direction of propagation of
the light string. To elucidate the physical origin of this force
we realize that expression (\ref{zetaq}) for the radiation
pressure force can be written as $F_{RP}=\hbar
k\cdot(I/\hbar\omega)\cdot\sigma$, where $\hbar k$ is the momentum
transferred to the electron along the z-axis per absorbed photon,
$(I/\hbar\omega)$ is the photon flux (number of photons per unit
area per unit time), and $\sigma=(ke^2\gamma/\omega
m\epsilon_0)/(\gamma^2+\omega^2)$ is the cross-section for inverse
Bremsstrahlung \cite{FeiFle75}. The force that causes the
electron-ion separation along the direction of propagation of the
light string may therefore be interpreted as the radiation
pressure force that accompanies inverse Bremsstrahlung in which
the electrons absorb photons in the field of the ions.

Our physical picture of the EMP generation is now as follows: The
light string of duration $t_p\approx 10$ fs generates an initial
dipole moment along the z-axis in the plasma channel which
subsequently oscillates at the plasma frequency
$\omega_{pl}=\sqrt{N_e e^2/m\epsilon_0}$ \cite{TonLan29,Jackson},
and radiates an EMP radially outwards from the channel (see
below). For electron densities $N_e\approx 10^{15}$ cm$^{-3}$
typical of light string induced plasmas, we have
$\omega_{pl}=2\times 10^{12}$ rads$^{-1}$ with corresponding EMP
wavelength $\lambda_{EMP}=2\pi c/\omega_{pl}\approx 1$ mm. Some
other physical considerations are in order regarding this picture.
In particular, in addition to the longitudinal electron-ion
displacement $\xi_0$ induced by the light string the electrons
also attain temperatures of the order of a few electron-volts
\cite{Vid00} due to the oscillatory force $-e\vec{E}_L$ acting on
them along the x-axis in the field of the light string.
Physically, electron-electron collisions will cause the electron
plasma, initially described by a non-equilibrium electron
distribution, to relax towards an Maxwellian distribution which as
a whole executes plasma oscillations along the z-axis
\cite{BonKoc}. We do not expect plasma oscillations along the
x-axis as the diameter of the plasma channel ($\approx 100$
$\mu$m) is smaller than the EMP wavelength. The ionized
electron-ion pairs can also undergo radiative recombination hence
producing high frequency EM fields in addition to the relatively
low frequency EMP of interest here.

To calculate the plasma-electron oscillations of the channel for
times after the light string has passed we use the classic
approach of Tonks and Langmuir \cite{TonLan29}: Then Eq.
(\ref{divE}) for the self-consistent electric field resulting from
the charge separation becomes $\partial E_z/\partial z=
(eN_e/\epsilon_0)(\partial\xi/\partial z)$, with solution
$E_z=(eN_e/\epsilon_0)\xi$. The longitudinal force on each
electron is then $-eE_z=m\partial\xi^2/\partial t^2$, which yields
the following equation of motion for the electron-ion displacement
\begin{equation}
\frac{\partial^2\xi}{\partial t^2} + \omega_{pl}^2(t)\xi = 0  ,
\label{plasmaosc}
\end{equation}
where $\omega_{pl}(t)=\sqrt{N_e(t)e^2/m\epsilon_0}$ as before.
Since the plasma density varies on a length scale $L$ of
centimeters which is much larger than the anticipated EMP
wavelength $\lambda_{EMP}\approx$ 1 mm, we hereafter treat the
plasma density, and hence the plasma frequency as constant over
the plasma channel. Equation (\ref{plasmaosc}) is to be solved
with the initial condition $\xi(0)=\xi_0$ from Eq. (\ref{xi0}):
This assumes that the initial electron-ion separation is not
significantly degraded by thermalizing collisions, meaning that we
are taking an upper bound for $\xi(0)$. For long times well after
the light string has passed, the electron density decays due to
radiative recombination as $\partial N_e/\partial t=-aN_e^2$
\cite{Tzortzakis:2000}, which has the analytic solution
$N_e(t)=N_e(0)/(1+atN_e(0))$. The initial value of the electron
density can be calculated by integrating Eq. (\ref{Neq}), but here
we use typical value of $N_e(0)=10^{15}$ cm$^{-3}$ obtained from
the full numerical solutions
\cite{Braun:1995,Nibbering:1996,Brodeur:1997,Mlejnek:1998,Mlejnek:1999}.

To obtain an approximate solution for the EMP due to the
oscillating electron plasma we write the longitudinal current as
$\vec{J}(\vec{r},t)=-eN_e(\partial\xi/\partial t)\hat{z}$.
Furthermore, since the plasma channel radius $r_0\approx 50$
$\mu$m is smaller than the EMP wavelength $\lambda_{EMP}\approx
1$ mm, and the plasma channel length $L>>\lambda_{EMP}$, we
approximate the induced current in the channel as a line source
\begin{equation}
\vec{J}(\vec{r},t)\approx -eN_e(t)\dot\xi(t)\delta(x)\delta(y)(\pi
r_0^2)\hat{z} , \label{curr}
\end{equation}
where the dot signifies a time derivative.  This current
oscillating along the z-axis will radiate a radially symmetric EMP
$\vec{E}_{rad}(\vec{r},t)=\hat zE_{rad}(r,t)$, independent of z
and dependent only on the radial distance r, that propagates
through air and radially away from the plasma channel, and obeys
Maxwell wave equation
\begin{equation}
\left (\nabla^2 - \frac{1}{c^2}\frac{\partial^2}{\partial t^2}
\right )\vec{E}_{rad}(\vec{r},t) =
\mu_0\frac{\partial\vec{J}}{\partial t} . \label{WavEq}
\end{equation}
It is important in this model that the EMP radiates into the air
surrounding the plasma channel as radiation at the plasma
frequency $\omega_{pl}$ has zero group velocity and cannot
therefore propagate in the plasma channel. Equation (\ref{WavEq})
along with the current in Eq. (\ref{curr}) can be solved using the
Green's function for the wave equation \cite{MW}
\begin{equation}
E_{rad}(r,t) = \frac{e\mu_0N_e(t)r_0^2}{4} \int_{-\infty}^\infty
dz \frac{\ddot\xi(t-\sqrt{r^2+z^2}/c)}{\sqrt{r^2+z^2}}
,\label{Erad}
\end{equation}
where we have taken the density $N_e(t)$ outside the integral based
on the fact that it decays on a time scale of nanoseconds, slow
compared to the plasma oscillations. Equation (\ref{Erad}) shows
that even though we have not included damping of the plasma
oscillation the radiated EMP is limited in duration by the time it
takes the electron plasma to recombine, which is on the order of a
few nanoseconds
\cite{Schillinger:99,Tzortzakis:99,Tzortzakis:2000,Vid00,Proulx:2000}:
In contrast, Landau damping \cite{Jackson} of the plasma
oscillations for our parameters occurred on a much longer time
scale.

\begin{figure} \epsfxsize 3 in \epsfbox{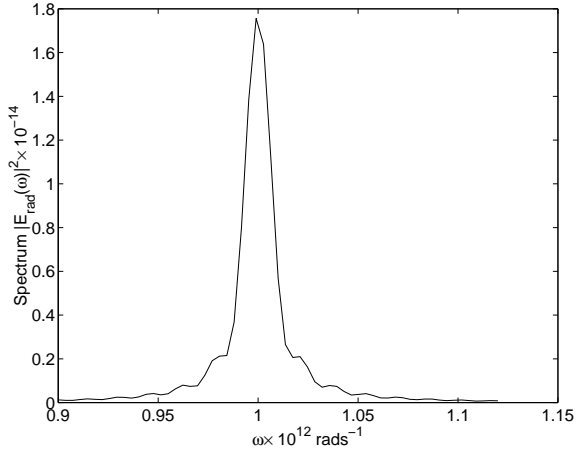}
\caption{Spectrum of the EMP.} \label{Fig:one}
\end{figure}

To illustrate the features of the EMP we have solved Eqs.
(\ref{plasmaosc}) and (\ref{Erad}) numerically with the initial
condition (\ref{xi0}).  Here we chose the light string parameters
$I_{0}=2\times 10^{13}$ W/cm$^{2}, t_p=10$ fs, an initial
electron density $N_e(0)= 10^{15}$ cm$^{-3}$, and $a=5\times
10^{-7}$ cm$^3$/s \cite{Mlejnek:1998}. We find that the radiated
EMP has a peak field strength of $|E_{max}|\approx 2$ kV/m at an
observation distance $r=1$ mm from the channel. The spectrum of
the EMP is shown in Fig. (\ref{Fig:one}), and the peak of the
spectrum is at $\omega_{peak}\approx 10^{12}$ rad s$^{-1}$, which
lies in the far IR region, which is consistent with the plasma
frequency for a density around $N_e\approx 10^{15}$ cm$^{-3}$. The
Hertzian width of the EMP spectrum is around $\Delta\nu\approx
3\times 10^{9}$ s$^{-1}$, which is consistent with the inverse
lifetime due to radiative electron recombination $aN_e(0)\approx 
10^9$ s$^{-1}$. Thus, the peak EMP frequency serves as a measure
of the plasma density, and the width of the spectrum is a measure
of the decay time of the electron plasma. This suggests that the
EMP from light string induced plasma could serve as a useful
diagnostic.

In summary, we have shown that the EMP from a light string induced
plasma channel can be understood as due to the establishment of a
dipole moment in the channel by the radiation pressure force of
the light string, which subsequently oscillates at the plasma
frequency and radiates. Having established the basic physics, in
future work we shall explore the details of the evolution of the
plasma channel using the full kinetic theory
\cite{ScoBinKoc92,KreBorBon99}, and in particular how the initial
dipole moment dephases before radiating. This raises the prospect
that the EMP field strength may be enhanced by creating $M$
parallel plasma channels \cite{Mlejnek:1999} within the EMP
wavelength, then the EMP intensity should be enhanced by $M^{2}$.
\vspace{0.5cm}

\noindent We are grateful to Profs. R. Binder and N. Bloembergen
at the Optical Sciences Center, Prof. M. Bonitz of the University
of Rostock, and Prof. S. Koch of the University of Marburg for
useful discussions regarding this work. This work is sponsored by
the U.S. Air Force Office of Scientific Research (AFOSR), under
grants AFOSR-F49620-00-1-002 and AFOSR-F49620-00-1-0190.

\end{document}